\documentclass[preprint,superscriptaddress,amsmath,amssymb,]{revtex4-1}

\usepackage[latin9]{inputenc}
\usepackage{amsmath}
\usepackage{graphicx}
\usepackage{ulem}

\makeatletter

\usepackage{epstopdf}
\usepackage{array}
\usepackage{mathrsfs}
\usepackage{dcolumn}
\usepackage{bm}
\usepackage{lineno}

\newcolumntype{L}[1]{>{\raggedright\let\newline\\\arraybackslash\hspace{0pt}}m{#1}}
\newcolumntype{C}[1]{>{\centering\let\newline\\\arraybackslash\hspace{0pt}}m{#1}}
\newcolumntype{R}[1]{>{\raggedleft\let\newline\\\arraybackslash\hspace{0pt}}m{#1}}

\graphicspath{{../../Inkscape/}{../../Mathematica/}}

\newcommand{\Rxx}{$\rho_{xx}$}
\newcommand{\Rxy}{$\rho_{yx}$}

\usepackage{tikz}
\usepgflibrary{shadings}
\usepackage{graphicx}
\usepackage{epstopdf}
\usepackage{textcomp}
\usepackage{dcolumn}
\usepackage{bm}
\usepackage{gensymb}
\usepackage{upgreek}
\usepackage{braket}

\begin{document}

\raggedbottom

\title{Structural tuning magnetism and topology in a magnetic topological insulator} 

\author{Christopher Eckberg}
\email{These two authors contributed equally}
\affiliation{Fibertek Inc., Herndon, Virginia 20171, USA}
\affiliation{DEVCOM Army Research Laboratory, Adelphi, Maryland 20783, USA}
\affiliation{DEVCOM Army Research Laboratory, Playa Vista, California 90094, USA}
\affiliation{Department of Electrical and Computer Engineering, University of California, Los Angeles, California 90095, USA}

\author{Gang Qiu}
\email{These two authors contributed equally}
\affiliation{Department of Electrical and Computer Engineering, University of California, Los Angeles, California 90095, USA}

\author{Tao Qu}
\affiliation{Department of Electrical and Computer Engineering, University of California, Los Angeles, California 90095, USA}

\author{Sohee Kwon}
\affiliation{Department of Electrical and Computer Engineering, University of California, Riverside, CA, 92521, US}

\author{Yuhang Liu}
\affiliation{Department of Electrical and Computer Engineering, University of California, Riverside, CA, 92521, US}

\author{Lixuan Tai}
\affiliation{Department of Electrical and Computer Engineering, University of California, Los Angeles, California 90095, USA}

\author{David Graf}
\affiliation{National High Magnetic Field Laboratory, Florida State University, Tallahassee, Florida, 32310, USA.} 

\author{Su Kong Chong}
\affiliation{Department of Electrical and Computer Engineering, University of California, Los Angeles, California 90095, USA}

\author{Peng Zhang}
\affiliation{Department of Electrical and Computer Engineering, University of California, Los Angeles, California 90095, USA}

\author{Kin L. Wong}
\affiliation{Department of Electrical and Computer Engineering, University of California, Los Angeles, California 90095, USA}

\author{Roger K. Lake}
\affiliation{Department of Electrical and Computer Engineering, University of California, Riverside, CA, 92521, US}

\author{Mahesh R. Neupane}
\affiliation{DEVCOM Army Research Laboratory, Adelphi, Maryland 20783, USA}
\affiliation{Department of Electrical and Computer Engineering, University of California, Riverside, CA, 92521, US}
\affiliation{Materials Science and Engineering Program, University of California, Riverside, CA, 92521, US}

\author{Kang L. Wang}
\email{wangkl@ucla.edu}
\affiliation{Department of Electrical and Computer Engineering, University of California, Los Angeles, California 90095, USA}

\date{\today}

\maketitle

\textbf{To date, the most widely-studied quantum anomalous Hall insulator (QAHI) platform is achieved by dilute doping of magnetic ions into thin films of the alloyed tetradymite topological insulator (TI) (Bi$_{1-x}$Sb$_x$)$_2$Te$_3$ (BST) \cite{chang2013experimental, kou2014scale, chang2015high,checkelsky2014trajectory}. In these films, long-range magnetic ordering of the transition metal substituants opens an exchange gap $\Delta$ in the topological surface states, stabilizing spin-polarized, dissipationless edge channels with a nonzero Chern number $\mathcal{C}$. The long-range ordering of the spatially separated magnetic ions is itself mediated by electronic states in the host TI, leading to a sophisticated feedback between magnetic and electronic properties. Here we present a study of the electronic and magnetic response of a BST-based QAHI system to structural tuning via hydrostatic pressure. We identify a systematic closure of the topological gap under compressive strain accompanied by a simultaneous enhancement in the magnetic ordering strength. Combining these experimental results with first-principle calculations we identify structural deformation as a strong tuning parameter to traverse a rich topological phase space and modify magnetism in the magnetically doped BST system.} 

Time-reversal invariant $\mathbb{Z}_2$ TIs feature gapless edge and surface states protected by time-reversal symmetry. Due to this time-reversal symmetry requirement, electronic band structures of $\mathbb{Z}_2$ TIs respond strongly to magnetic perturbation \cite{yu2010quantized, zhang2013topology, liu2016quantum}. This relationship is notably exemplified in the realization of the quantum anomalous Hall effect in magnetic TI systems. In QAHIs, long-range magnetic order gaps the otherwise mass-less topological surface state. When the chemical potential is positioned within the exchange gap, the 2D density of states vanishes, and a chiral edge state represents the lone channel for electrical transport. The resulting phase is characterized by a combination of long-range magnetic order, quantized Hall conductivity, and vanishing longitudinal resistance; all of which persist in the absence of an applied magnetic field. The promise of technologically significant phenomena in QAHI materials including dissipationless, non-reciprocal electrical transport \cite{yasuda2020large, mahoney2017zero}, quantized magneto-electric dynamics \cite{essin2009magnetoelectric, qi2008topological}, and exotic quasiparticle excitations \cite{zeng2018quantum,wang2015chiral} to name a few, has stimulated a tremendous research effort in QAHI systems and magnetic topological matter in general.

Though recent discoveries have notably expanded the landscape of known QAHI hosts \cite{deng2020quantum, liu2020robust, serlin2020intrinsic}, the most mature and widely studied QAHI platform is the Cr substituted \newline (Bi$_{1-x-y}$Sb$_x$Cr$_y$)$_2$Te$_3$ (CBST) system grown using molecular beam epitaxy (MBE). Due to the greatly reduced spin-orbit coupling strength of Cr compared with the Bi/Sb atoms it substitutes, in quantized CBST the concentration of magnetic dopants must be left relatively dilute lest they promote a topological phase transition to a trivial insulating state \cite{zhang2013topology,chang2014}. The dilute magnetism and disordered, dual-doped crystal structure of CBST imbues an unfortunate fragility onto the quantum anomalous Hall effect in CBST at elevated temperatures \cite{pan2020probing,li2016origin,lee2015imaging,tokura2019magnetic}. This fragility of the quantum anomalous Hall state presents a major hurdle limiting the technical applicability of these compounds. Operational temperatures may be improved to a degree by varying dopant concentrations and profiles \cite{mogi2015magnetic,ou2018enhancing}. However, in practice chemical optimization of these materials is a delicate and imperfect process, as chemical composition simultaneously impacts the positioning of the chemical potential, electronic band structure, disorder profile, and magnetic ordering strength. A cleaner tuning parameter to more directly engineer CBST band structures is therefore essential to improve CBST QAHI operating temperatures.

Here we report the magnetic and electronic evolution of CBST in response to a continuous deformation of the crystal lattice via hydrostatic pressure. Pressure dependent experiments were performed on gated Hall bar devices at pressures up to 1.6 GPa and temperatures as low as 20 mK. Our experiments demonstrate the electronic and magnetic properties of CBST to be highly responsive to strain, with lattice compression both suppressing the QAHI phase and enhancing the magnetic order. First-principle calculations confirm these effects emerge from a structural driven evolution of the CBST band structure, and indicate a rich topological phase space may be addressed through the application of even larger pressures. Together, these experimental and theoretical results demonstrate crystal strain as an effective tuning parameter to selectively modify the low energy electronic structure in BST based magnetic topological matter, establishing structural engineering as a viable pathway to control critical material properties in the future.

Experiments are conducted on 6 quintuple layer (QL) thick MBE grown CBST films with a magnetic Curie temperature $T_C$ of roughly 20 K. Data are presented for three different photolithographically defined Hall bar devices, labelled P1, P2, and P3. Samples P2 and P3 were fabricated simultaneously from the same wafer in field-effect transistor geometries, where an approximately 20 nm thick HfO$_x$ layer serves as the gate dielectric. These two devices display virtually identical behaviors over a wide range of temperature and magnetic field \cite{Supplement}, and, in the following, their properties will frequently be compared directly. P1 meanwhile was fabricated from a separate wafer. The impact of pressure on the topological transport signatures and magnetism of these devices was studied using a piston cell equipped for electrical transport experiments. During experiment, P1 was measured in a dilution refrigerator while P2 and P3 were primarily studied in a $^3$He sorption cryostat. Taken together, data gathered on these different devices span nearly three decades in temperature ranging in regime from $k_bT<<\Delta$ to $k_bT \approx T_C$. 

\begin{figure*}
    \centering
    \includegraphics[width=0.97\textwidth]{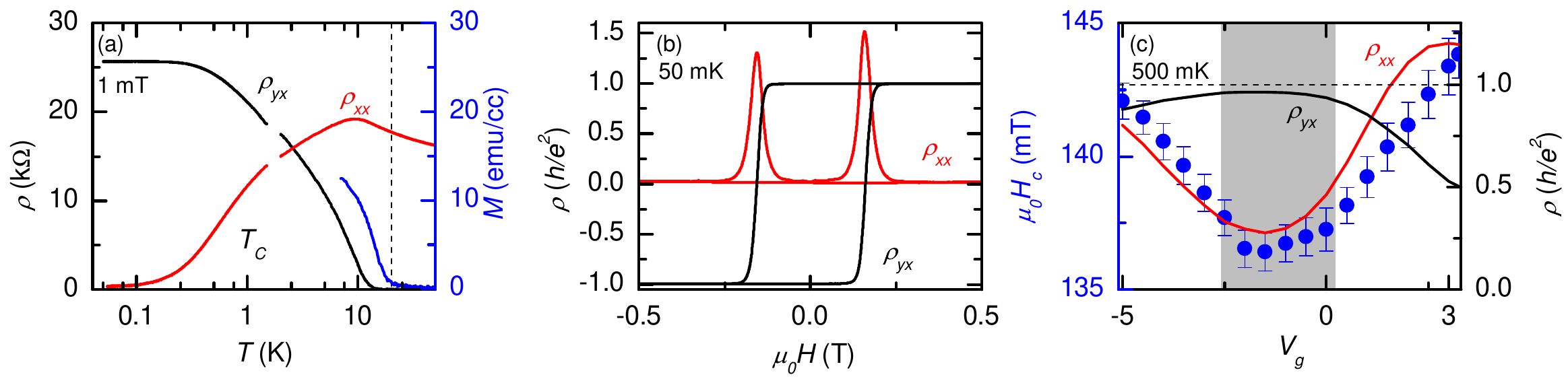}
    \caption{{\bf Summary of magnetic and electrical properties in a QAHI.} 
    (a) Temperature dependent \Rxx~(red) and \Rxy~ (black) are presented for a QAHI device P2. Temperature dependent sample magnetization acquired using a piece of unpatterned film is included for comparison (blue). All data was collected at a magnetic field of 1 mT. Curie temperature denoted in the figure is determined by Arrott analysis \cite{Supplement}. (b) Field dependent \Rxx~ and \Rxy~ data collected at 50mK demonstrating well-quantized transport behavior. (c) Gate dependent \Rxx~ (red), \Rxy~ (black), and magnetic coercive field $\mu_0H_c$ (blue) at a temperature of 500 mK are displayed. In these data, $\mu_0H_c$ was determined by the location of the zero crossing points of the Hall bar's two $\rho_{yx}$ channels. The error bars, meanwhile, represent the standard deviation of the four separate transitions measured ($i.e.$ up-down and down-up transitions in each channel). Grey shading denotes the center of the emerging topological gap.  
    }
    \label{fig:Figure1}
\end{figure*}

We begin by presenting the ambient pressure properties of device P2 (Fig. \ref{fig:Figure1}). At $T_C$, the systems develops a magnetization when subjected to a small external field. When cooled to lower temperatures, this magnetic order manifests a rapidly increasing anomalous Hall signal, and at temperatures well below $T_C$ \Rxx~begins to rapidly decrease as seen in Fig. \ref{fig:Figure1} (a). At dilution refrigerator conditions, \Rxx~approaches zero while \Rxy~approaches the quantized value of $h/e^2\approx 25.8$ k$\Omega$. Field dependent hysteresis loops in a dilution refrigerator environment are presented in Fig. \ref{fig:Figure1} (b). In these data, transitions between \Rxy~plateaus accompany the switching of the magnetic order in the system between the down and up states and mark a topological transition between $\mathcal{C}=1$ and $\mathcal{C}=-1$. The $\rho_{xx}$ peaks and $\rho_{yx}$ zero crossings observed in the magnetic hysteresis loops occur at the magnetic coercive field $\mu_0 H_c$ and correspond with an $M_z=0$ condition (Fig. \ref{fig:Figure1} (b)). 

Figure \ref{fig:Figure1} (c) demonstrates the gate response of device P2 at a temperature of $T=500$ mK. At this relatively elevated temperature thermal excitations into dissipative states precludes the high quality quantization observed in the dilution cooled regime. Nevertheless, a clear $\rho_{xx}$ ($\rho_{yx}$) minimum (maximum) is observed at an optimized gate voltage of roughly $-1.5$ V; indicative of the incipient QAHI phase. The magnetic coercive field $\mu_0H_c$, a rough avatar for the magnetic ordering strength, is also measured as a function of the gate voltage. We observe an enhanced magnetic order when the system is driven away from the charge neutral point. Such an enhancement in magnetism with the addition of carriers into the system has been previously reported, and is commonly attributed to itinerant carrier mediated RKKY interactions strengthening the coupling between Cr-ions \cite{zhang2014electrically, kou2015magnetic, kou2013interplay}. In a narrow range near $V_g^c$, however, $\mu_0H_c$ exhibits very little if any response to the changing gate voltage (gray regime in Fig. \ref{fig:Figure1} (c)). In this region the carrier concentration is minimized, and the Cr-Cr magnetic coupling is sustained by the van Vleck mechanism \cite{yu2010quantized,li2015experimental,ji2022thickness}.

\begin{figure*}
    \centering
    \includegraphics[width=0.97\textwidth]{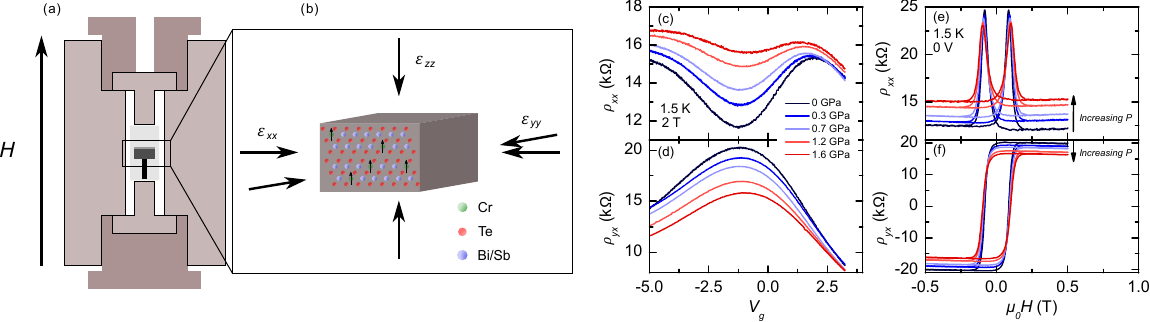}
    \caption{{\bf Pressure dependent \Rxx~and \Rxy~in a QAHI device.} (a) Schematic of pressure dependent experiment, depicting the geometry of the pressure cell used and the orientation of the sample plane and the external magnetic field. (b) Cartooned depiction of hydrostatic pressure effect on CBST unit cell, emphasizing the roughly isotropic compression anticipated in these experiments. (c,d) Gate dependent measurements of \Rxx~(c) and \Rxy~(d) collected at 1.5 K and 2 T. (e,f) Field dependent \Rxx~(e) and \Rxy~(f) collected at 1.5 K and a $V_g$ of 0 V. For data presented in panels (c-f), 0 GPa data were collected on device P3 while all data at non-zero pressure were collected from sample P2. 
    }
    \label{fig:Figure2}
\end{figure*}

Following ambient pressure characterization, QAHI devices were loaded into a piston pressure cell (Fig. \ref{fig:Figure2}(a)) and studied at hydrostatic pressures up to 1.6 GPa (Fig. \ref{fig:Figure2} (b)). Comparison to the related Sb$_2$Te$_3$ system suggests a nearly isotropic compression of roughly 1\% may be expected in our CBST films at the maximal pressure \cite{zhu2013superconductivity, souza2012high}, though clamping by the more rigid GaAs substrate may slightly mute the lattice compression experienced by our thin film devices \cite{al2004empirical,souza2012high}. Despite the modest compression that may be anticipated in these experiments, our QAHI devices are nevertheless quite responsive to lattice tuning in the range of pressure studied. Figure \ref{fig:Figure2} presents a summary of basic transport data collected at 1.5 K, indicating that the system trends away from quantized transport with shrinking unit cell size. This is evidenced by an increase in $\rho_{xx}$ and concomitant reduction in $\rho_{yx}$ seen in both gate voltage traces (Fig. \ref{fig:Figure2} (c), \ref{fig:Figure2} (d)) and field hysteresis loops (Fig. \ref{fig:Figure2} (e), \ref{fig:Figure2} (f)). Despite the trend away from quantized transport, a $\rho_{yx}$ ($\rho_{xx}$) maxima (minima) is still seen near $V_g^c$ at all pressures. That the gate cannot recover the same degree of transport quantization at all pressures suggests the flow away from idealized QAHI behavior is due to a pressure driven modification of the electronic band structure rather than a rigid shift of the chemical potential, as the latter scenario could possibly be compensated for by sweeping out carriers using the gate. In fact, the observation of a consistent $V_g^c$ at all pressures suggests any shifting of the Fermi energy within the pressure range explored is minimal.

\begin{figure*}
    \centering
    \includegraphics[width=0.97\textwidth]{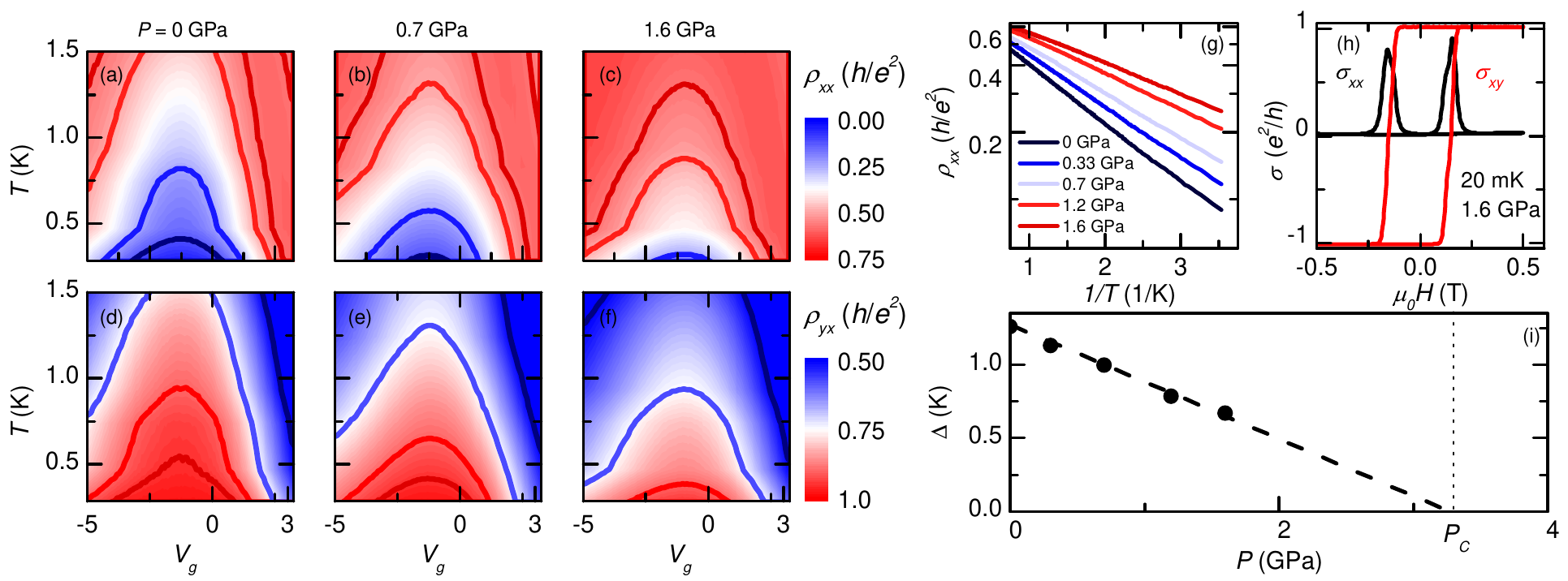}
    \caption{{\bf Evolution of topological gap with increasing pressure.} (a-f)
    Temperature and gate dependent \Rxx~ and \Rxy~ are presented at a constant field of $\mu_0H=2$ T, and pressures of 0 GPa (a,d), 0.7 GPa (b,e), and 1.6 GPa (c,f). Contour lines are included at values of 0.125, 0.25, 0.5, and 0.558 $h/e^2$ in \Rxx~color plots, and at values of 0.5, 0.75, 0.9, and 0.95 in \Rxy. (g) Logarithm of longitudinal resistance values presented versus 1/T. The linearity of the curves when presented in this fashion confirm thermally activated transport behavior of the form $\rho_{xx}(T)\propto \exp(-\Delta/2k_BT)$.  (h) Magnetic hysteresis loop of sample P1 collected at a pressure of 1.6 GPa and temperature of 20 mK confirming the persistence of high quality quantization at dilution temperatures and high pressures. (i) Pressure dependence of topological gap determined by temperature dependencies presented in (g). Linear extrapolation to zero predicts a critical pressure $P_c$ of approximately 3.3 GPa.}
    \label{fig:Figure3}
\end{figure*}

Comparison of the temperature and voltage dependent $\rho_{xx}$ and $\rho_{yx}$ at pressures of 0, 0.7, and 1.6 GPa are shown as two-dimensional color plots in Figs. \ref{fig:Figure3} (a-f), providing a qualitative visualization of a closing topological gap with increasing pressure. By fitting the temperature dependent $\rho_{xx}$ at $V_g=0$ V and $\mu_0H=2$ T to an Arrhenius model (Fig. \ref{fig:Figure3} g) the value of this gap is quantified at all pressures. The pressure dependence of the topological gap is summarized in Fig. \ref{fig:Figure3} (i), indicating the gap remains intact, but decreases in a linear fashion by almost exactly a factor of 2 from 1.2 K to 0.6 K over the pressure range explored. Extrapolating the linear trend to zero suggests a critical pressure $P_C$ of approximately 3.3 GPa, at which point we anticipate a topological phase transition away from the QAHI state to occur. To demonstrate that the QAHI phase does persist to the highest pressures measured in this study, in Fig. \ref{fig:Figure3} (h) we present data collected on another device, sample P1, at a temperature of 20 mK and pressure of 1.6 GPa. At these conditions, we still observe conductivity values within 3\% of the quantized expectation of $e^2/h$, confirming that 1.6 GPa is insufficient to drive these samples out of the QAHI ground state.

\begin{figure}
    \centering
    \includegraphics[width=0.47\textwidth]{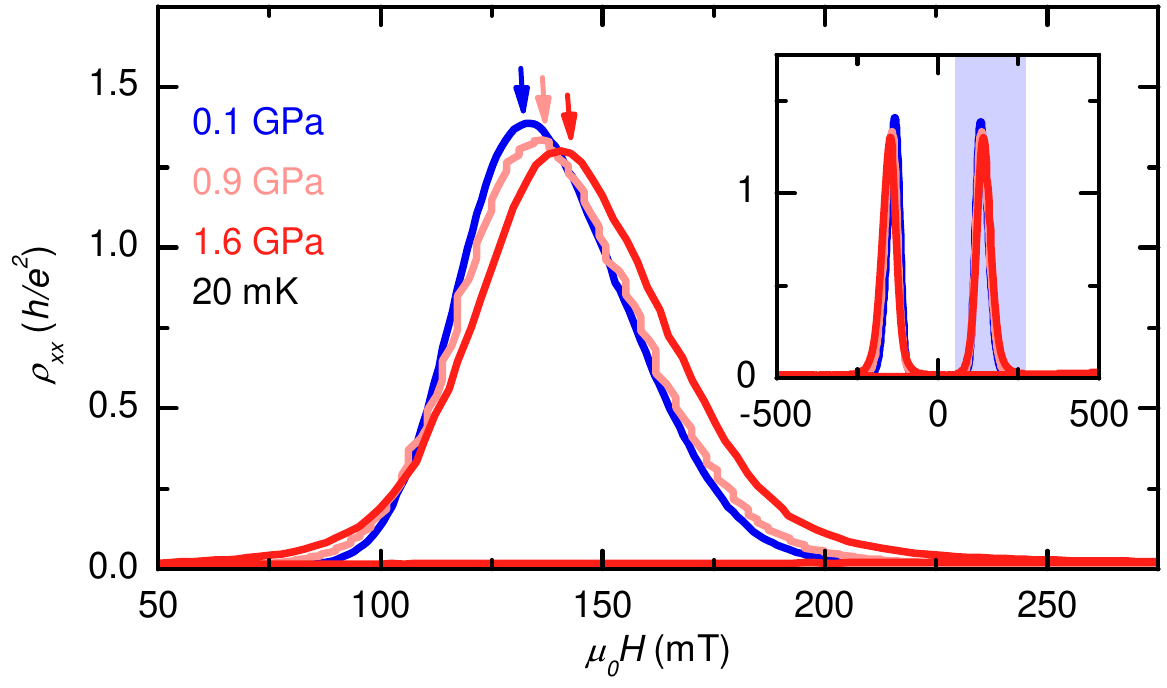}
    \caption{{\bf Evolution of magnetism under hydrostatic pressure.} 
    Pressure dependent evolution of \Rxx~hysteresis loops are shown for sample P1 at 20 mK and pressures of 0.1, 0.9, and 1.6 GPa. The inset displays the full magnetic hysteresis loops, while the main figure is zoomed to the region near $\mu_0H_c$ shaded in blue in the inset. 
    }
    \label{fig:Figure4}
\end{figure}

Intriguingly, while the transport gap closes in pressurized QAHI samples, we observe a pressure driven \textit{enhancement} in the magnetic order as demonstrated by an increasing coercive field. This enhancement is visible in Fig. \ref{fig:Figure4}, where the field dependent \Rxx~hysteresis loops collected from device P1 at a temperature of 20 mK are presented. The \Rxx~peaks are clearly pushed to higher fields with increasing pressure, reflecting the growth in $\mu_0H_c$ from a value of 133 mT at 0.1 GPa to a notably larger 144 mT at 1.6 GPa. At the 20 mK temperature where these data are collected $k_bT<<\Delta$. Consequently, the strength of magnetic interactions dependent upon itinerant charge carriers are vanishingly weak and this enhanced magnetic order is presumably sustained through the van Vleck mechanism. In addition to the magnetic enhancement observed at 20 mK, an increasing coercive field under pressure is also observed in devices P2/P3 between 280 mK and 15 K, demonstrating this effect to be consistent between samples and persistent over a wide temperature range \cite{Supplement}. Gate-dependent data collected in P2/P3 demonstrate the magnetic enhancement can be maximized by gate-tuning towards the valence band; indicating that pressure likely also enhances the hole-mediated RKKY interaction in a manner consistent with previous reports in more traditional dilute magnetic semiconductors \cite{csontos2005pressure}. 

The strengthened magnetic ordering indicates that it is unlikely that the pressure driven reduction in the transport gap is a consequence of a reduced exchange field at the Dirac surface states. Alternative sources that may account for the suppressed gap include increasing surface hybridization or occupation of delocalized, dissipative states. To understand what effects are dominant in our system, we perform first principle band structure calculations for a 6 QL thick slab of CBST host compound Sb$_2$Te$_3$. As Bi primarily functions as a counter-dopant in CBST \cite{zhang2011band}, and Cr $d$-electrons do not contribute to the density of states at the Fermi energy \cite{yu2010quantized}, Sb$_2$Te$_3$ calculations capture the principal details of the CBST band structure \cite{ji2022thickness} while notably excluding the material magnetism and resulting exchange gap. Thus, in the presented calculations, trends in the surface hybridization are observed directly and are not obfuscated by magnetic gapping of the surface band structure. Following previous example \cite{zhu2013superconductivity, zhang2011pressure}, pressure is simulated by isotropically compressing the boundaries of the computational unit and allowing all interior atoms to relax to their lowest energy configuration. 

\begin{figure*}
    \centering
    \includegraphics[width=0.95\textwidth]{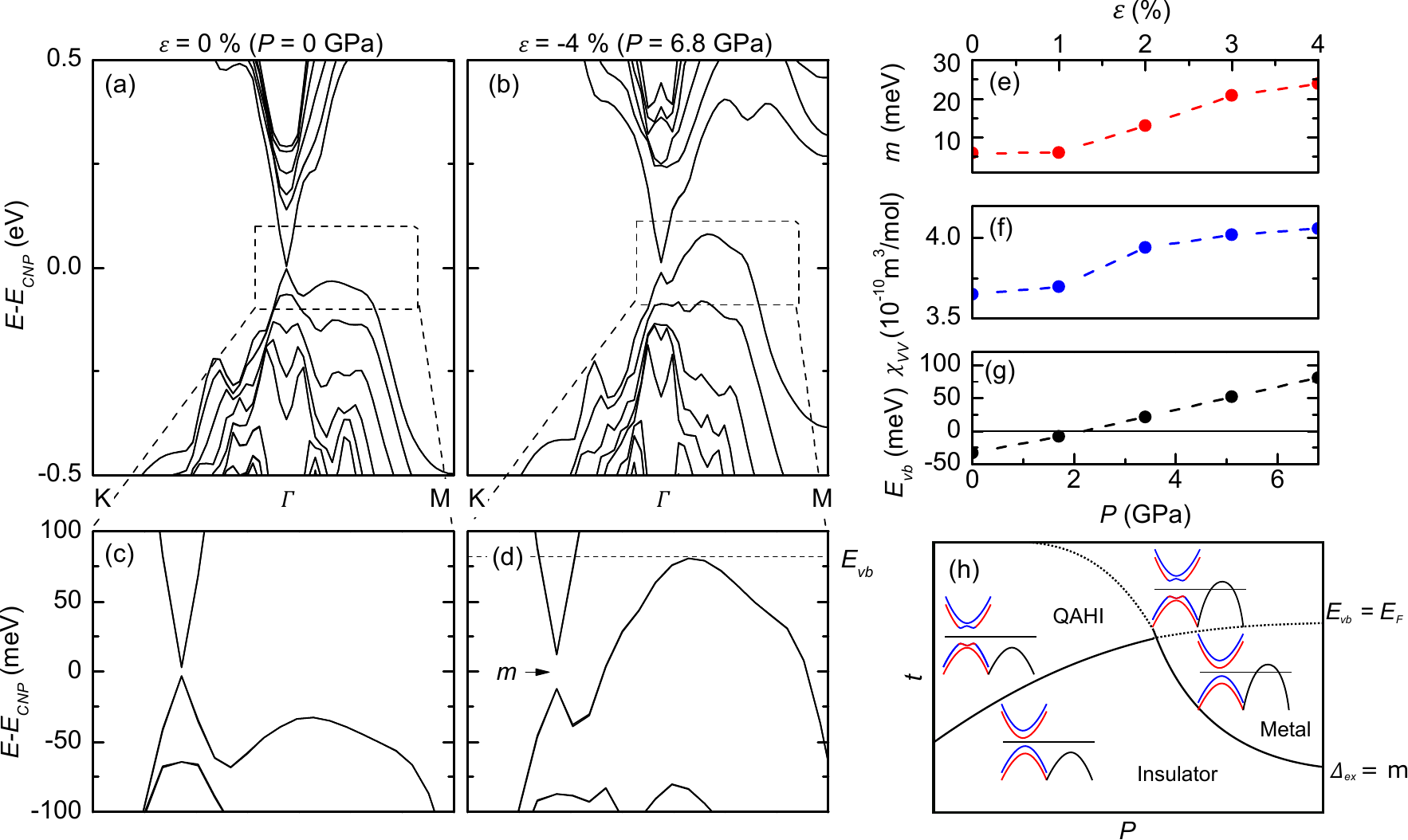}
    \caption{{\bf Electronic evolution with hydrostatic pressure.} 
    (a-d) Calculated electronic band structure of a 6 QL thick Sb$_2$Te$_3$ slab with isotropic lattice compressions of 0\% (a) and -4\%. To simplify their comparison, the calculated band structures are shifted vertically so that the center of the surface band gap ($E_{CNP}$) rather than the calculated $E_F$ is positioned at zero energy. (b). Zoomed band structures near $E_F$ and $\Gamma$ are shown in (c) (0\%) and (d) (4\%), highlighting the pressure dependencies of the hybridization gap $m$ and the energy of the valence band valley $E_{vb}$. The locations of the zoomed regions are marked in (a) and (b) by boxes. (e-g) Pressure dependence of $m$ (e), $\chi_{VV}$ (f), and $E_{vb}$ (g) are presented with dashed lines as a guide for the eye. (h) Proposed zero-temperature topological phase diagram for the CBST system as a function of pressure $P$ and film thickness $t$. Cartooned band structures representative of the electronic ground state are shown in the four distinct regions in this topological phase space. In these simplified cartoons, the red and blue lines represent the spin split surface bands, while the bulk valence band is presented in black.
    }
    \label{fig:Figure5}
\end{figure*}

Band structure calculations were performed in 1\% strain increments between 0\% and 4\%. Calculations at 0\% and -4\% compressive strains are presented in Fig. \ref{fig:Figure5} (a-d). Consistent with previous reports, we find that increasing pressure widens the direct, bulk band gap at $\Gamma$, while simultaneously raising the energy of the valence band valley $E_{vb}$ between $\Gamma$ and $M$ \cite{zhu2013superconductivity}. At the computational thickness of 6 QL, we observe a small hybridization gap $m$ in the surface bands even in the zero-compression limit (Fig. \ref{fig:Figure5} (e)). This feature is amplified as the unit cell size is decreased, indicating increasingly pronounced hybridization between top and bottom surfaces of the TI under pressure. Finally, using the calculated band structures, we extract the van Vleck susceptibility $\chi_{VV}$ according to the relationship:
\begin{equation}
\chi_{VV} = \frac{1}{N}\sum_k \sum_{E_{nk}<\mu<E_{mk}}4 \mu_0 \mu_B^2 \frac{\bra{nk}\hat{S_z}\ket{mk}\bra{mk}\hat{S_z}\ket{nk}}{E_{mk}-E_{nk}}
\label{eq:VV}
\end{equation} 
\noindent Here $\mu_0$ is the vacuum permeability, $\mu_B$ is the Bohr magneton, $\hat{S_z}$  is the spin operator. $n$, $\ket{nk}$, and $E_{nk}$ represent the band index, wave function and eigenvalue of $n$th band in valence bands at momentum $k$, while $m$, $\ket{mk}$ and $E_{mk}$ correspond to conduction band states. Based upon these calculations we observe a clear though modest enhancement in $\chi_{VV}$ with increasing pressure (Fig. \ref{fig:Figure5} (f)). This enhancement emanates primarily from states near $\Gamma$, suggesting it is born from increasing mixing between the inverted Sb $p_{1z}^-$ and Te $p_{2z}^-$ states in the compressed lattice. 

The above observations have significant implications to the QAH state. The pressure driven enhancement in $m$ biases the system towards a trivial insulator phase, with the electronic phase transition occurring once the magnitude of $m$ grows larger than that of the exchange gap $\Delta_{ex}$. Meanwhile, once the energy of the valence band valley exceeds the Fermi energy (i.e. $E_{vb}>0$) delocalized states in the valence band will populate down to lowest temperatures and the system will behave as a metal. In the case when $E_{vb}>0$ and $\Delta_{ex}>m$ carrier conduction in the chiral channels will relax internally through the dissipative valence band states, precluding transport quantization. The calculated evolutions of $m$, $\chi_{VV}$, and $E_{vb}$ with increasing pressure (Fig. \ref{fig:Figure5} (e-g)) therefore imply a coalescence of metallic, insulating, and QAHI electronic ground states in the pressure tuned CBST system. Based upon these calculated band structures and the known evolutions of the hybridization gap, bulk state quantum confinement, and exchange energy with decreasing CBST thickness \cite{ji2022thickness,pan2020probing,feng2016thickness}, we present a proposed topological phase diagram in layer thickness and pressure dependent parameter space in Fig. \ref{fig:Figure5}(h). These phase boundaries could be further adjusted by tuning along other axes such as external magnetic field \cite{ji2022thickness, pan2020probing} or applied gate voltage. We will also note that the QAHI/insulator phase boundary presented here assumes that the hybridization gap is more responsive to pressure than the exchange gap, an assumption supported by the relatively weak pressure effect on $\chi_{vv}$ (Fig. \ref{fig:Figure5} (f)) compared with $m$ (Fig. \ref{fig:Figure5} (e)).  It is possible, however, that the exchange gap may feature its own pressure dependence, altering the trajectory of the $\Delta_{ex}=m$ boundary.

Having discussed the calculated pressure dependent band structure for this system, we now consider how these calculations comport with the experimentally determined results. Notably, our band structure calculations unambiguously indicate a trend away from the QAHI state in compressed tetradymite TIs, consistent with the experimental reality. While the calculations indicate that the electronic state beyond $P_c$ may be either trivially insulating or metallic depending upon the details of the material, in our samples we believe the topological phase transition occurring at $P_c$ to be towards the metallic regime. We come to this conclusion through the observations of reduced \Rxx~ at temperatures above $T_C$, as well as reductions in the \Rxx~peaks at $\mu_0H_c$ with increasing pressure; both of which suggest an increasing density of states near the Fermi energy. Meanwhile, the enhanced $\chi_{vv}$ observed in calculation captures the increasing magnetic ordering strength observed in our pressurized QAHI films.   

To conclude, these results establish lattice deformation as an effective, clean tuning parameter for modifying the electronic and magnetic properties of alloyed QAHI materials. Though a significant material response is observed in the pressure range explored in this study, we believe increasing pressure may evoke even more dramatic electronic and magnetic responses. Crucially, $P_c$ is well below the 9+ GPa threshold at which a structural phase transition from rhombehedral to monoclinic crystallographic point groups has been previously reported in tetradymite TI systems \cite{vilaplana2011,zhu2013superconductivity,kirshenbaum2013pressure}, indicating future experiments may explore a significantly larger pressure range without concern of interference from additional structural phases. Finally, on the basis of these results, we propose that tensile strain, as opposed to its compressive counterpart studied here, may present an exciting tuning parameter to explore in future efforts to enhance QAHI behavior.

\section{\label{sec:Methods}Methods}

\subsection{Material Growth}

All CBST films were grown in an ultra-high vacuum, Perkin-Elmer molecular beam epitaxy (MBE) system. Epi-ready semi-insulating GaAs (111)B substrates were used for the growth. Before growth, the substrates were loaded into the MBE chamber and pre-annealed at the temperature of 630 \textdegree C in a Te-rich environment in order to desorb the oxide on the surface. During growth, the substrate was kept at 190 \textdegree C. High-purity Bi, Sb, Cr and Te sources were evaporated simultaneously from standard Knudsen cells. The growth process was monitored by the reflection high-energy electron diffraction (RHEED) \textit{in-situ}, and the digital RHEED images were captured using a KSA400 system built by K-space Associates, Inc. Sharp and streaky lines in the RHEED pattern indicate good epitaxial crystalline quality.

\subsection{High pressure experiments}

Pressure was applied using a standard piston pressure-cell. To fit within the active area of the pressure cell, single devices were cut from pre-patterned wafers to dimensions of less than 3.0 mm, and were fixed to a fiber optic using epoxy to orient the sample within the pressure cell. Thin platinum wires were attached by hand to the contact pads of the device under test with silver paint. A small ruby chip was fixed to the tip of the fiber optic, which was used to calibrate the pressure at room temperature and again at low temperature. A PTFE cup was filled with Daphne 7575 oil and fixed in place over the sample platform so that the device was surrounded by the hydrostatic fluid.  Once assembled, the cell was placed in a hydraulic press where a piston fed through a hole in the threaded top screw of the cell was used to add pressure. When the appropriate pressure was reached, the top screw was clamped, locking in the pressure.

Transport measurements were collected using a low-frequency ($<10$ Hz) lockin technique with $ac$ excitations of 10 nA. Gate swept data display a small hysteresis based upon the gate history. To compensate for this effect and ensure consistency all data presented were taken during sweeps from $+3.25$ to $-5$ V.

Cryogenic sample environments for ambient pressure experiments were maintained using a Quantum Design Physical Property Measurement System equipped with a dilution refrigerator insert. High pressure measurements were conducted in high magnetic field cells SCM-1 and SCM-2 at the National High Magnetic Field Laboratory in Tallahassee. SCM-1 is equipped with a dilution refrigerator cryogenic environment while SCM-2 was operated with pure $^3$He cooled using a sorption pump. Both SCM-1 and SCM-2 are equipped with 18 T superconducting magnets. To compensate for the remnant field of the superconducting magnet, the field dependent data were calibrated using a Hall sensor. Additionally, to account for the magnetoresistance of the SCM-2 thermometers, the temperatures used in Figs. \ref{fig:Figure3} and \ref{fig:Figure4} were calibrated using the strong temperature dependence of the QAHI material itself. For details of the field and temperature calibration processes please refer to the Supplemental Information \cite{Supplement}.

\subsection{First principle calculations}

We perform first-principles calculations as implemented in the Vienna Ab Initio Simulation Package (VASP) \cite{kresse1996efficient}.The Perdew, Burke, Ernzerhof (PBE) form of the generalized gradient approximation is used as the exchange-correlation functional \cite{perdew1996generalized}. The computational cell employed is constructed from six QL Sb$_2$Te$_3$ slabs stacked with a 40 \AA~thick vacuum region. We apply an energy cutoff of 500 eV and a 8x8x1 $\Gamma$-centered $k$-grid to optimize cell structure and atomic positions. The optimized lattice constant of the slab is $a=b=4.3307$ \AA~and $c=31.09$ \AA. Tri-axial compressive strains between -4.0\% and 0.0\% are applied by shrinking the perimeter of the Sb$_2$Te$_3$ slab. At each strain, atomic positions inside the unit cell are allowed to relax in all directions. Spin-orbit coupling is included during the charge density relaxation for electronic band structures.

Based upon the band structure, van Vleck susceptibilities were calculated according to Eq. \ref{eq:VVmethod} \cite{yu2010quantized}.

\begin{equation}
\chi_{VV} = \frac{1}{N}\sum_k \sum_{E_{nk}<\mu<E_{mk}}4 \mu_0 \mu_B^2 \frac{\bra{nk}\hat{S_z}\ket{mk}\bra{mk}\hat{S_z}\ket{nk}}{E_{mk}-E_{nk}}
\label{eq:VVmethod}
\end{equation}  

\noindent where, as described in the main text, $\mu_0$ is the vacuum permeability, $\mu_B$ is the Bohr magneton, $\hat{S_z}$  is the spin operator. $n$, $\ket{nk}$, and $E_{nk}$ represent the band index, wave function and eigenvalue of $n$th band in valence bands at momentum $k$, while $m$, $\ket{mk}$ and $E_{mk}$ correspond to conduction band states. The susceptibility averages over $k$ points in the first Brillouin zone, where $N$ is the number of $k$ points. We focus on the $k$ symmetric lines ($K-\Gamma-M$) around the Dirac point which make the susceptibility calculation feasible.

The wave functions $\ket{nk}$ is a spinor, with both spin up and spin down components due to spin-orbit coupling, as shown in Eq. \ref{eq:spinor}.

\begin{equation}
\ket{nk} = 
\begin{pmatrix}
\psi_{nk}^\uparrow (\textbf{r})\\
\psi_{nk}^\downarrow (\textbf{r})
\end{pmatrix}
\label{eq:spinor}
\end{equation}  

\noindent Here $\psi_{nk}^\uparrow (\textbf{r})$ and $\psi_{nk}^\downarrow (\textbf{r})$ are spin up and down components of a real space wave function. The real space wave functions are found through summing over plane-wave vectors $\textbf{G}$ and their associated plane-wave coefficients. The wavefunctions are read from WAVECAR files, produced by VASP \cite{feenstra2013low}, with a cut-off plane-wave energy of 500 eV.

\section{\label{sec:Acknowledge}Acknowledgments}

C.E. is an employee of Fibertek, Inc. and performs in support of Contract No.W15P7T19D0038, Delivery Order W911-QX-20-F-0023. The views expressed are those of the authors and do not reflect the official policy or position of the Department of Defense or the US government. The identification of any commercial product or tradename does not imply endorsement or recommendation by Fibertek Inc. This work was supported by the NSF under Grants No. 1936383 and No. 2040737, the U.S. Army Research Office MURI program under Grants No. W911NF-20-2-0166 and No. W911NF-16-1-0472.  A portion of this work was performed at the National High Magnetic Field Laboratory, which is supported by the National Science Foundation Cooperative Agreement No. DMR-1644779 and the State of Florida. This work was supported in part by the DEVCOM Army Research Laboratory (ARL) Research Associateship Program (RAP) Cooperative Agreement(CA) W911NF-16-2-0008. This work used the Extreme Science and Engineering Discovery Environment (XSEDE) \cite{towns2014xsede}, which is supported by National Science Foundation Grant No. ACI-1548562 and allocation ID TG-DMR130081.

\section{\label{sec:Contributions}Author Contributions}

C.E., G.Q., L.T., and K.L.W. conceived of and designed the experiments. QAHI materials were grown by P.Z. and L.T.. Devices were fabricated by G.Q., S.K.C., and K.W.. Pressure dependent transport measurements were performed by C.E., G.Q., and D.G.. First principle calculations were performed by S.K., Y.L., and M.N., and T.Q. calculated the van Vleck susceptibilities. C.E., G.Q., and K.L.W. wrote the manuscript with contributions from all authors.

\section{\label{sec:Data Availability}Data Availability}

The data represented in the figures are available with the online version of this paper. All other data that supports the plots within this paper and other findings of this study are available from the corresponding author upon reasonable request.

\section{\label{sec:Competing Interests}Competing Interests}

The authors declare no competing interests.

\bibliographystyle{apsrev4-1}

\bibliography{MTIPressure_refs}

\pagebreak
\clearpage

\pagebreak
\clearpage

\pagebreak
\clearpage

\pagebreak
\clearpage

\pagebreak
\clearpage

\end{document}